# Neural Replicator Analysis of the Genus *Flavivirus*

Alexandr A. Ezhov

State Research Center of Russian Federation
Troitsk Institute for Innovation and Fusion Research,
108840, Troitsk, Moscow, Russia
ezhov@triniti.ru

## Abstract

The results of applying neural replicator analysis (NRA) to the genomes of viruses belonging to the genus *Flavivirus* are presented. It is shown that the viral genomes considered in this study can be placed in five different cells of the viral genome table. Some of these cells appear for the first time and are characterized by 9-periodicity of WS-encoded genomic sequences. It is noteworthy that Japanese encephalitis viral strains and Zika viral strains occupy not one, but two common cells of this table. We also present the results of the NRA of Zika viral strains and suggest that the earliest strain in Asia is an Indian strain that spread from Africa (Uganda) to the East. The fine structure of the sets of Japanese encephalitis viral strains is presented and it is shown that their generally accepted genotypes 1 and 3 can be clearly divided into two subgenotypes. It is also shown that probably not Indonesian, but Indian strains of this virus can be considered the earliest known strains that further evolved and spread in Asian countries.

## Introduction

Interest in flaviviruses, in addition to their importance as the cause of many dangerous diseases and outbreaks, is also associated with their oncolytic properties. Moore, the first to investigate oncolytic viruses using developed rodent cancer models, used an *in vivo* tumor model to convincingly demonstrate that an oncolytic virus, in this case Russian Far East encephalitis virus, can selectively target and kill cancer cells in living animals, reporting their initial results in 1949. She found that in some cases, mouse sarcoma 180 can be completely destroyed [1].

Recently, a new nomenclature for Zika viruses has been proposed in [2]. This once again showed the existence of different views on the delicate classification of viral strains. One new reason for this was the discovery of a new strain of Zika in India. As the authors of [2] stated:

> “The recent sample from India, also from a human host (OK054351_India_28 July 2021), clustered with an old Malaysia sequence but with an unexpected high divergence. This finding warrants further investigation to understand the reason for the accumulation of such high divergence” [2].

In this paper, we demonstrate that NRA [3,4], which provides a new insight into the classification of the viral genomes, clearly shows the close similarity of these Indian and Malaysian strains. This can be seen as further confirmation of the usefulness of the NRA. In addition, here we apply NRA to the study of the entire *Flaviviruses* genus of the *Flaviviridae* family and argue for the existence of finer structures of the Zika and Japanese encephalitis viruses, which likely provide new information about the origin and evolution of these two viruses. The structure of the article is as follows. In the first section, we present the results of applying NRA to the study of the genus *Flavivirus* and show that their strains can be placed in five separate cells of the viral genome table presented in [4]. It was revealed that viral genomes of Japanese encephalitis and Zika strains occupy two common cells of the table. In the second section, we present NRA results for Zika viral strains and suggest that the earliest strain in Asia is an Indian strain that spread from Africa (Uganda) to the East. The third section presents the fine structure of the Japanese encephalitis viral strain sets and shows that their generally accepted genotypes 1 and 3 can be clearly divided into two subgenotypes. We also make the assumption that possibly not Indonesian, but Indian strains of this virus can be considered the earliest known, which subsequently spread and evolved to the countries of the East. In the Conclusion section, we present a discussion of the results obtained using NRA.

## 1. Neural Replicator Analysis of the genus *Flavivirus*

The application of NRA to viruses of the *Flavivirus* genus showed that they can be placed in five cells of the table of natural classification of viral genomes introduced in [4][A3]. Some preliminary remarks about the approximations made here are necessary. The flavivirus genomes are formed by a linear strand of RNA, but for simplicity and generality, we assume that the NRA considers them to be circular, as was done for the truly circular DNA strands of viruses considered in previous works [3,4]. This approximation seems to be quite precise, since the maximum length of the sliding window (the maximum number of neurons in the replicator network) is chosen as before equal to $K$=30. Since the length of flavivirus RNA sequences exceeds 10 000, this approximation introduces an error of only about 0.3% into the analysis. We also characterize the absence of replicators for $K$ larger than 30 with the abbreviation NoR (No Replicators), which means that they really do not exist for networks of all considered sizes of $K \leq 30$ .

*Tick-borne viruses*

The most unusual cell (9, NoR), characterized by the presence of the same common 9-period motif for WS-encoded genomes and the absence of replicators for KM-encoded ones, is colonized by tick-borne mammalian host viruses (Table 1, Fig. 1). ). Most of them also have transmission patterns with additional length periods of 3, 11, 13. In addition, two viruses with an unknown vector belonging to the Rio Bravo group of viruses - Montana myotic leukoencephalitis virus and Rio Bravo virus - and the Yutiapa strain JG- 128 (having many additional motif periods) also have 9-period motives (Table 1, Fig. 1). Therefore, the NRA definitely suggests that only ticks can be their vectors. Two exceptions to this trait of tick-borne viruses are the mammalian Powassan tick virus strains, the LB strain, and the Deer tick strains, whose replicators have unique 6-periodic motifs: as well as additional 2-, 3-, and 13-period ones. (Fig. 1, Table 2). The Powassan virus exists in North America and was also discovered in the Far East of the USSR in 1978. The genomes of these viruses can be conditionally assigned to a cell (6, NoR).

Table 1. Flaviviruses belonging to the cell (9, NoR). Below, the state of neuron -1 is denoted as 0.

| Virus | Abbrev | Accession | 9-period motif | Additional periods, 3,11,13 |
|---|---|---|---|---|
| Tick-borne encephalitis - European | TBEV-Eur | U27495.1 | 101010100 | 3: 110, 100 |
| Tick-borne encephalitis - Siberian | TBEV-Sib | L40361.3 | 101010100 | 3: 110, 100 |
| Tick-borne encephalitis – Far East | TBEV-FE | KF889893 | 101010110<br>101010100 | 3: 110, 100 |
| Kyasanur forest disease virus | KFDV | AY323490.1 | 101010100 | 3: 110, 100 |
| Alkhurma viral strain 1176 | AHFV | AF331718.1 | 101010100 | 3: 110, 100 |
| Omsk hemorrhagic fever virus | OHFV | AY193805 | 101010100 | 3: 110, 100 |
| Royal Farm virus | RFV | DQ235149 | 101010100 | 3: 110, 100 |
| Greek goat encephalitis virus | GGEV | DQ235153.1 | 101010100 | 3: 110, 100<br>13: 1010101010100 |
| Turkish sheep encephalitis virus | TSEV | DQ235151.1 | 101010100 | 3: 110, 100;<br>13: 1010101010100 |
| Spanish sheep encephalitis virus | LIV-Spain | DQ235152.1 | 101010100 | 11: 10101010110 |
| Gadgets Gully virus | GGYV | DQ235145 | 101010100 | 11: 10101010110 |
| Louping ill virus | LIV | Y07863.1 | 101010100 | 11: 10101010100 |
| Langat viral strain TP21 | LGTV | AF253419 | 101010100 | 2: 10; 3: 110, 100<br>11: 10101010100 |
| *Montana myotic leukoencephalitis virus | MMLV | AJ299445 | 101010100 | 11: 10101010110 |
| *Rio Bravo virus | RBV | AF144692.1 | 101010100 | 11: 10101010110 |
| *Jutiapa viral strain JG-128 | JUTV | KJ469371 | 111011110 | 3: 100<br>5:10000,10110,10010<br>7: 1010110 |

Table 2. Flaviviruses belonging to the cell (6, NoR).

| Virus | Abbrev | Accession | 6-period motifs | Additional periods, 2, 3,13 |
|---|---|---|---|---|
| Powassan viral strain B | POWV | L06436.1 | 101000 | 2: 10; 3: 110, 100 |
| Deer tick viral strain ctb30 | DTV | AF311056 | 100000, 101100 | 13: 1001011010010 |

The other cell (3, NoR) is occupied by the genomes of tick-borne viruses of seabirds (Table 3).

Table 3 Flaviviruses belonging to the cell (3, NoR).

| Virus | Abbrev | Accession | 3-period motifs | Additional periods, 2, 3,13 |
|---|---|---|---|---|
| Meaban virus | MEAV | DQ235144.1 | 110, 100 | - |
| Saumarez Reef virus | SREV | DQ235150.1 | 110, 100 | - |
| Tyuleniy virus | TYUV | KF815939.1 | 110, 100 | |

*Mosquito-borne viruses*

However, most of the viruses that inhabit this cell are carried by mosquitoes.

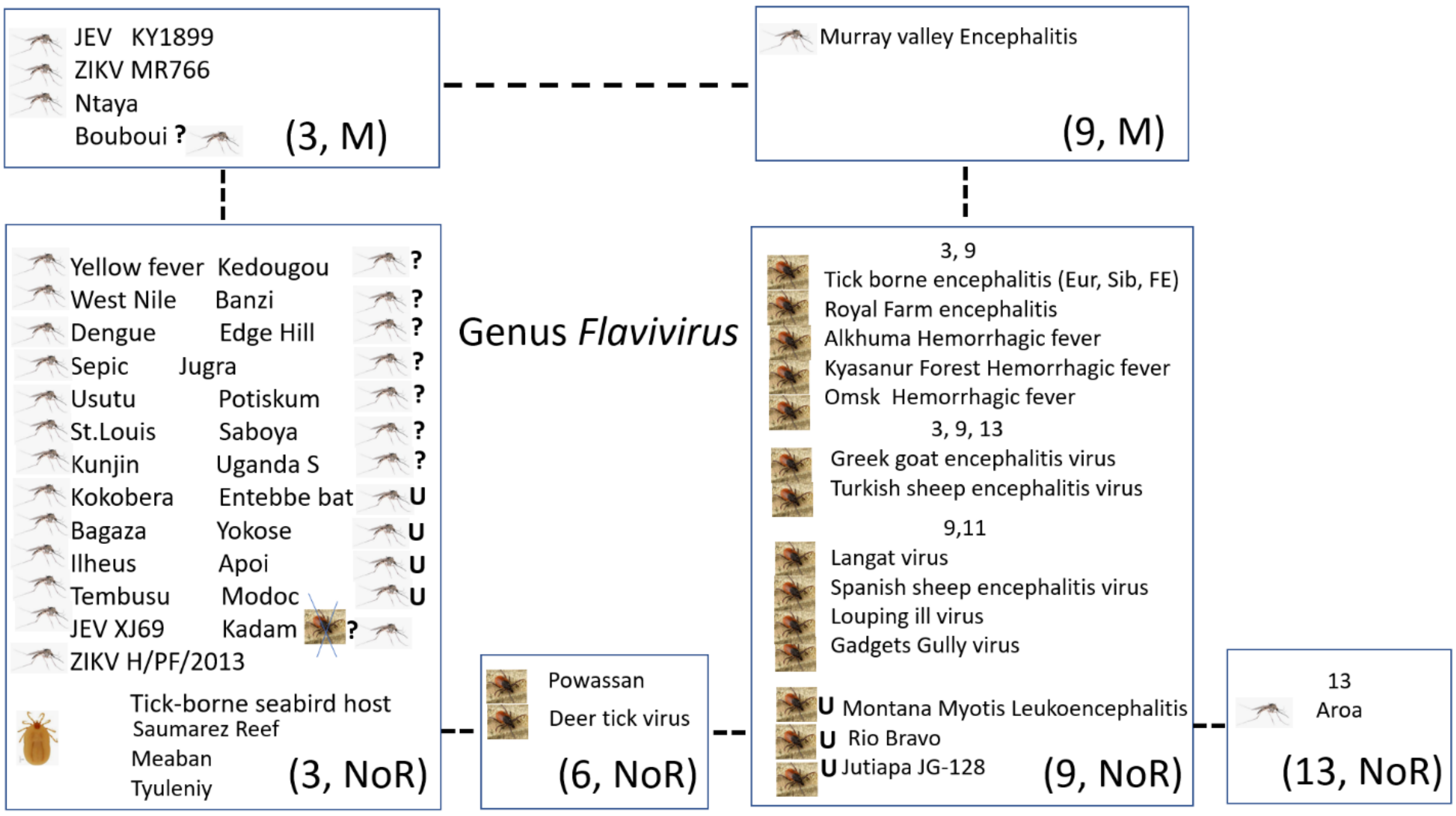

**Fig. 1.** Six cells – (3, NoR), (6, NoR), (9, NoR), (13, NoR), (3,M) and (9,M), occupied by tick and mosquito strains of flaviviruses. The question mark denotes strains whose vector is characterized as "*probably*", "U" – strains with an "*unknown vector*". Zika viral (ZIKV) and Japanese encephalitis viral (JEV) strains occupy two cells, (3, NoR) and (3, M). NRAs reliably differentiate tick-borne and mosquito-borne viruses and in most cases confirm hypothetical vectors. Cell (3, NoR) is presumably occupied by mosquito-borne viruses, as well as only specific tick-borne viruses of seabirds. Cells (3, M), (9, M), (13, NoR) are occupied only by mosquito-borne viruses. Cell (9, NoR) is occupied by tick-borne viruses without exception. Accession numbers of all strains are presented in Tables 4,5,6.

Table 4. Mosquito-borne flaviviruses belonging to the cell (9, NoR).

| Virus | Abbrev | Accession | motif | Additional periods |
|---|---|---|---|---|
| Zika viral strain H/PF/2013 | ZIKV | KJ776791.2 | 3:100 | |
| Japanese encephalitis viral strain XJ69 | JEV | EU880214.1 | 3:100 | |
| West Nile virus | WNV | M12294.2 | 3:100 | |
| Kunjin virus | KUNV | D00246.1 | 3:100 | |
| Yellow fever virus | YFV | X03700.1 | 3:100 | |
| Sepic virus | SEPV | DQ837642.1 | 3:100 | |
| St Louis encephalitis virus | SLEV | DQ525916.1 | 3:100 | |
| Usutu virus | USUV | AY453411.1 | 3:100 | |
| Kokobera virus | KOKV | AY632541.4 | 3:100 | |
| Bagaza virus | BAGV | AY632545.2 | 3:100 | |
| Ilheus virus | ILHV | AY632539.4 | 3:100 | |
| Tembusu virus | TMUV | JF895923.2 | 3:100 | |
| Dengue virus 1 | DENV-1 | U88536.1 | 3: mixture: 100, 010 | |
| Dengue virus 2 | DENV-2 | U87411.1 | 3: mixture: 100, 010 | 2:10 |
| Dengue virus 3 | DENV-3 | M93130.1 | 3: 100 | |
| Dengue virus 4 | DENV-4 | AF326573.1 | 3: 100 | |

A very exotic case is the mosquito-borne Aroa virus (AROAV, AY632536.4), whose WS-encoded replicators transmit patterns with a period of 13 (1001001001100). The NRA shows that many other viruses that are characterized as "probably mosquito-borne" also enter the cell (3, NoR).

Thus, the NRA supports the hypothesis that these viruses are carried by mosquitoes (Table 5, Fig. 1).

Table 5. Flaviviruses that are suspected to be carried by mosquitoes belong to the cell (3, NoR). Thus the NRA assumes that they are actually carried by mosquitoes.

| Virus | Abbrev | Accession | motif |
|---|---|---|---|
| Kedougou virus | KEDV | AY632540.2 | 3:100 |
| Banzi virus | BANV | DQ859056.1 | 3:100 |
| Edge Hill virus | EHV | DQ859060.1 | 3:100 |
| Jugra virus | JUGV | DQ859066.1 | 3:100 |
| Potiskum virus | POTV | DQ859067.1 | 3:100 |
| Saboya virus | SABV | DQ859062.1 | 3:100 |
| Uganda S virus | UGSV | DQ859065.1 | 3:100 |

Finally, some viruses with an unknown vector are also assigned to the cell (3, NoR). This supports the mosquito as such a vector.

Table 6. Flaviviruses placed in a cell (3, NoR) with an unknown vector are likely to be transmitted by mosquitoes according to the NRA.

| Virus | Abbrev | Accession | motif |
|---|---|---|---|
| Entebbe bat virus | ENTV | DQ837641.1 | 3:100 |
| Yokose virus | YOKV | AB114858.1 | 3:100 |
| Apoi virus | APOIV | AF160193.1 | 3:100 |
| Modoc | MODV | AJ242984.1 | 3:100 |

However, the "*probably tick-borne*" Kadam virus from Uganda (KADV, DQ235146) is directed by the NRA to cells colonized by mosquito-borne and tick-borne seabird host viruses, making the original hypothesis suspect. NRA analysis shows that many (but not all) strains of Zika virus and Japanese encephalitis virus belong to the cell (3, NoR). Some examples of these strains are shown in Table 4 and in Fig. 1.

The Ntaya virus (NTAV, JX236040.3) belongs to the (3,M) cell, where M denotes the presence of KM-encoded genome replicators with a generally *monotonous* (M) set of non-periodic transmitted patterns with gradually fading activity (a decreasing number of positive neuronal states). In addition, the suspicious Boubou virus (BANV, DQ859056.1) belongs to the (3, M) cell, and the NRA also confirms that it is transmitted by the mosquito. The exotic Murray Valley Encephalitis virus is alone at (9, M). As we will see below in a detailed examination of the strains of the Zika virus, as well as the Japanese encephalitis virus, they can massively populate the (3, M) cell (some of them - the ZIKV M766, AY632535.2 strain and the JEV KU1899, AY316157 strain are shown in Fig. 1).

## 2. Neural Replicator Analysis of Zika viral genomes

Two main genetic clusters of Zika viruses are called African and Asian genotypes, respectively [5]. The Asian genotype is subdivided into Asian and American (and Oceanic) lines. As noted in [2], there is still no classification system that reflects the true extent of Zika virus genetic variability. In 2022, the authors of [2] proposed to divide the Asian lineage into subgroups that have a basal Malaysian sequence and a recently discovered Indian sequence, and a subgroup that includes sequences from Southeast Asia, South Asia, and Micronesia (also the third subgroup contains sequences related to outbreak in Singapore). The American (and Oceanian) lines fall into three subgroups, associated with South America (and French Polynesia), Central America, the Caribbean, and North America. Neural Replicator Analysis of Zika viral strains, which, as in [2], examined their complete genomes (the first such analysis was described in 2007 [6]), gives interesting results that can be obtained with a closer look at the fine structure of Replicator Tables and which not only qualitatively correspond to early classification schemes, but also provide information on possible ways of distribution and evolution of viruses. As we mentioned above, the Zika virus (like the Japanese encephalitis virus, which will be discussed in the next section) may not have replicators for KM-encoded RNA, for all network sizes up to $K$=30 (such strains belong to the cell (3, NoR) ), or they may have replicators with monotonic sets of transmitted patterns (which implies random character of patterns stored in the Hopfield ancestor network) − the cell (3, M). Remarkably, in the second case, replicators appear only for networks whose sizes exceed the threshold value $K \geq K_{min}$=20! Thus, the Replicator Tables have the form shown in Fig. 2. In some rare cases, replicators do not exist for some intermediate values of $K$: $K_{min} \leq K \leq 30$, but in most cases this does not happen.

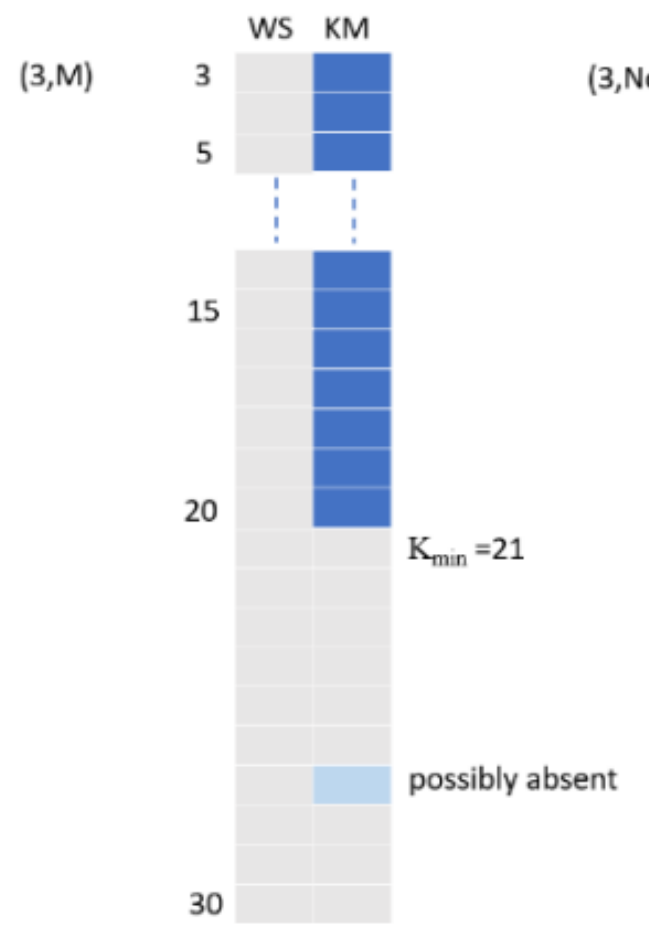

**Fig.2.** Replicator Tables of Zika strains belonging to the cell (3, M) – on the left and (3, NoR) – on the right. For WS-encoded genomes, replicators exist for all network sizes $K$ up to 30, and a periodicity of 3 is observed for transmitted patterns. For KM-encoded genomes, replicators appear only starting from $K=K_{min}$, where $20 \leq K_{min} \leq 30$.

*African genotype*

Zika virus was discovered in 1947 in Uganda and later found in Central and West African countries. The NRA results show that the first Uganda strain, MR766, has an extreme Replicator Table for which $K_{min}$ = 20 is the minimum value for all strains studied. The most distant West African Senegal Zika strains have higher $K_{min}$ values and are concentrated mainly in the cell (3, NoR) – Fig.3. Note, that two distinct clusters are formed, one near $K_{min}$ =22 and one in cell (3, NoR).

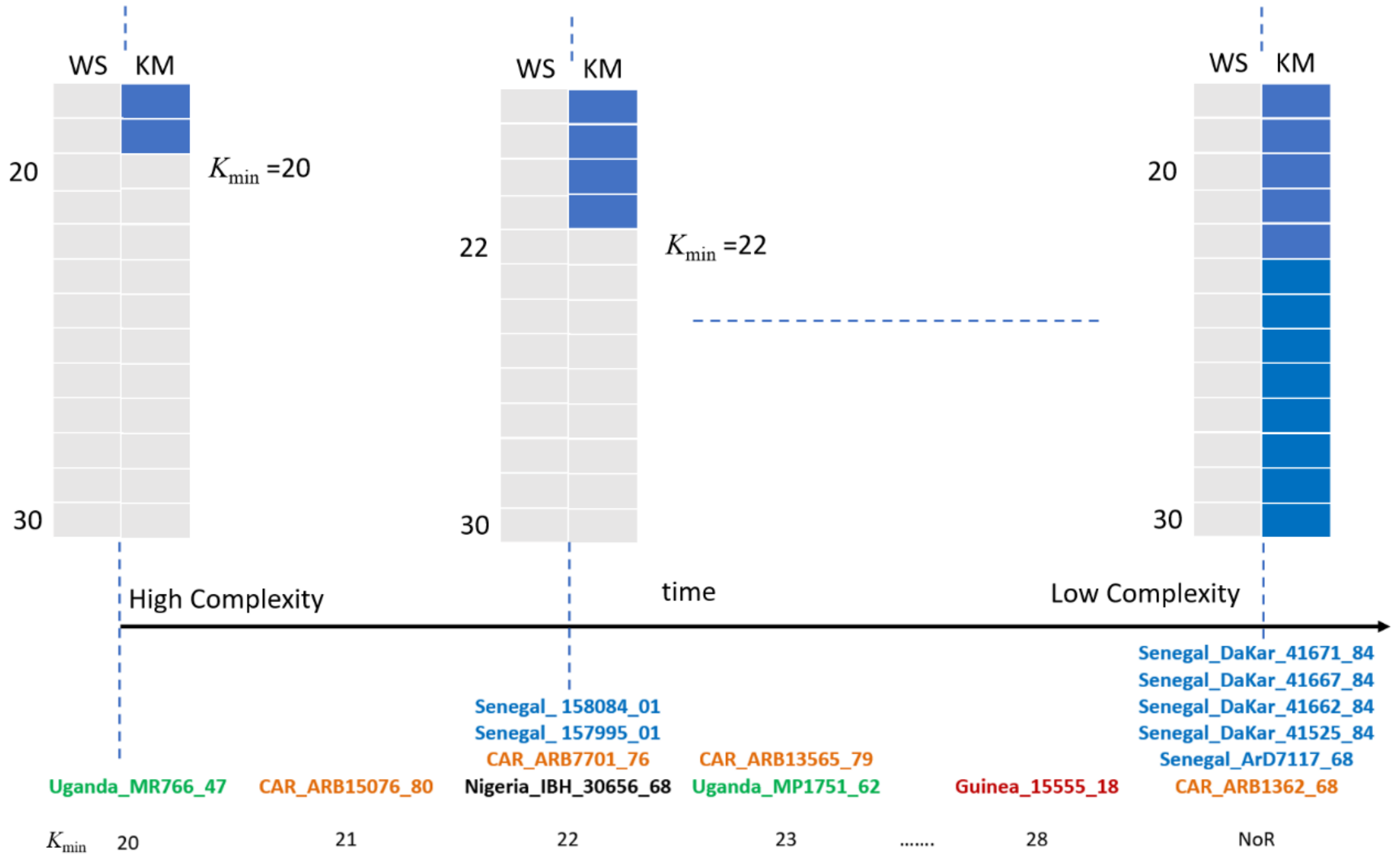

**Fig. 3.** Top: Example of African Zika Replicator Tables. The right RT corresponds to viruses that do not have replicators for KM-encoded genomes with a size less than or equal to $K$=30. The respective strains refer to the cell (3, NoR). Other RTs (left and middle) describe strains placed in a cell (3,M). Bottom: Distribution of African strains of the Zika virus according to the $K_{min}$ value, which defines the minimum replicator size generated from the original Hopfield neural network trained on KM-encoded sequences. The Uganda MR766 strain has the lowest $K_{min}$ value and is expected to have the highest complexity and an earlier origin than the other strains. The West African Senegalese viruses are assumed to be in cell (3, NoR) and considered to be more recent and less complex.

Most strains of Senegal virus are in the cell (3, NoR) – the right column in Fig. 3 (Table 7) . These strains are considered as emerged later than the Uganda strain, and have simpler Replication Tables without replicators for $K \leq 30$. On the other hand, Uganda strain has, in a sense, the most complex form, with minimum threshold for the appearance of replicators, consequently, with a maximum number of replicators of different sizes (Table 8). It can be assumed that NRA indicates that the evolution of the Zika virus is accompanied by a simplification of viral genomes. This assumption is justified when considering the Asian strains of the Zika virus, the distribution of which is shown in Fig.4.

Table 7. African Zika strains belonging to the cell (3, NoR) and presented in Fig.3

| Viral strain | Accession | Viral strain | Accession | Viral strain | | Accession |
|---|---|---|---|---|---|---|
| CAR_ARB1362_68 | KF383115.1 | Senegal_ArD7117_68 | KF383116 | Senegal_DaKar_41525_84 | | KU955591.1 |
| Senegal_DaKar_41662_84 | KU955592 | Senegal_DaKar_41667_84 | MF510857.1 | Senegal_DaKar_41671_84 | | KU955595 |

Table 8. Strains of the African Zika virus placed in cell (3,M) – Fig.3. The $K_{min}$ values correspond to the minimum size of the network processing KM-encoded genomes, and generating replicators are presented along with the corresponding sets of patterns transmitted by replicators. The $K_{max}$ values correspond to the size of the replicator network that has the maximum number of transmitted patterns.

| Zika virus strain | Accession | $K_{min}$ | Transmitted patterns for $K_{min}$ | $K_{max}$ | Transmitted patterns for $K_{max}$ |
|---|---|---|---|---|---|
| Uganda_MR766_47 | LC002520.1 | 20 | 00000000000000001111<br>00000000000000000111 | 23 | 00000000011111111111111<br>00000000000000011111111<br>00000000000000001111111<br>00000000000000000111111<br>00000000000000000001111<br>00000000000000000000111<br>00000000000000000000011 |
| CAR_ARB15076_80 | KF268949 | 21 | 000000000000000001111 | 23 | 00000000011111111111111<br>00000000000000001111111<br>00000000000000000111111<br>00000000000000000011111<br>00000000000000000001111<br>00000000000000000000111<br>00000000000000000000011 |
| Nigeria_IBH_30656_68 | KU963574.2 | 22 | 0000000000000000001111 | 30 | 000000000000000011111111111111<br>000000000000000000111111111111<br>000000000000000000000111111111<br>000000000000000000000011111111<br>000000000000000000000001111111 |
| CAR_ARB7701_76 | KF268950 | 22 | 0000000000000000001111<br>0000000000000000000111<br>0000000000000000000011 | 23 | 00000000000000000011111<br>00000000000000000001111<br>00000000000000000000111<br>00000000000000000000011 |
| Senegal_157995_01 | KF383118.1 | 22 | 1110000000000000000000 | 22 | 1110000000000000000000 |
| Senegal_158084_01 | KF383119 | 22 | 1111111111111100000000<br>1111111111111000000000<br>1111111100000000000000<br>1111000000000000000000<br>1110000000000000000000<br>1100000000000000000000 | 22 | 1111111111111100000000<br>1111111111111000000000<br>1111111100000000000000<br>1111000000000000000000<br>1110000000000000000000<br>1100000000000000000000 |
| Uganda_MP1751_62 | KY288905.1 | 23 | 00000000000000001111111<br>00000000000000000111111<br>00000000000000000011111<br>00000000000000000001111 | 23 | 00000000000000001111111<br>00000000000000000111111<br>00000000000000000011111<br>00000000000000000001111 |
| CAR_ARB13565_79 | KF268948 | 23 | 00000000000000000000111<br>00000000000000000000011 | 23 | 00000000000000000000111<br>00000000000000000000011 |
| Guinea_15555_18 | MN025403.1 | 28 | 0000000000000000000001111111 | 28 | 0000000000000000000001111111 |

*Asian genotype*

The NRA results are presented in Fig. 4 and in Tables 9, 10. It is noteworthy that the Malaysian and Indian strains, characterized in [2] as basal of the first subgroup of the Asian lineage, both have an extremal value of $K_{min}$ =21, which is far from the corresponding values of other strains of Asian genotype (Fig. 4). Remind that in [2] the authors stated that the recent sample from India, also from a human host (OK054351_India_28 July 2021), clustered with an old Malaysia sequence, but with an unexpected high divergence.

Note that, considering the forms of Replicator Tables, as well as the patterns trainsmitted by replicators, we see that these two strains are almost identical. Thus, NRA does not face the problem of high divergence as noted in [2], given that both approaches are based on the analysis of complete viral genomes and despite the difference of genome lengths (10 269 nt for Malaysian versus 10 748 nt for Indian strains).

The recent detection of strain MR766 in Uganda during the 2019 Madhya Pradesh Zika outbreak in India and its close similarity of RT ($K_{min}$ =20) to the RT of strain Ind_OK054351_21 ($K_{min}$ =21) may support the hypothesis that Zika virus was transmitted from Africa to India and Malaysia and then to other Asian countries. Note also the viral sequence found from a patient in Gujarat appears to be similar to the Malaysian strain.

The problem of interpretation of RTs obtained in NRA and the relationship of their form with the characteristics of the viral strains is, of course, quite complex. We have already suggested that a higher RT complexity (lower $K_{min}$) may indicate an earlier origin of virus. This is consistent with the form of the first MR766 strain found  Uganda.  Also, the Senegal strains mainly occupy the cell (3, NoR), which corresponds to the transmission of the virus from the East to West Africa. Finally, comparing Figures 3 and 4, we see that the Asian strains are also much more populated in cell (3, NoR) than the African strains. This is also consistent with the spread of African strains to Asia, Africa and Oceania.

Note that the 2016 WHO Reference Candidate strain (KX369547), isolated from the serum of a patient from French Polynesia in 2013 (PF13_25013-18), proposed as a representative of clinically significant viruses with a wide distribution [7] (this sequence, however, does not cover the entire length of the genome), as well as another reference genome (H/PF/201, 3KJ776791) proposed later in 2017 as the Zika virus reference [8], are located in the same cell (3, NoR), which can be considered populated by the latest strains of Zika viruses.

Many works have been devoted to comparing the properties of African and Asian strains. For example, in [9], when comparing Asian and African strains, the authors formulated a paradoxical situation where all human outbreaks and birth defects to date are associated exclusively with Asian Zika viruses, despite laboratory data indicating higher transmissibility and pathogenicity of African Zika viruses. Note however, that in 2019 African Uganda strain MR766 was found during India outbreak. The authors of [9] argued that the Thailand strain (Tai_sv0127_14) ($K_{min}$=24) was associated with more adverse outcomes than the F_Polynesia_2013 epidemic strain (cell (3, NoR)). This was also confirmed by the following publication [10], where only the Ugandan strain MR766 was also mentioned as causing more embryonic damage than the Asian strains. Thus, some relationship between the forms of RT and the pathogenicity of viral strains may indeed exist, but this hypothetical relationship is subject to further study.

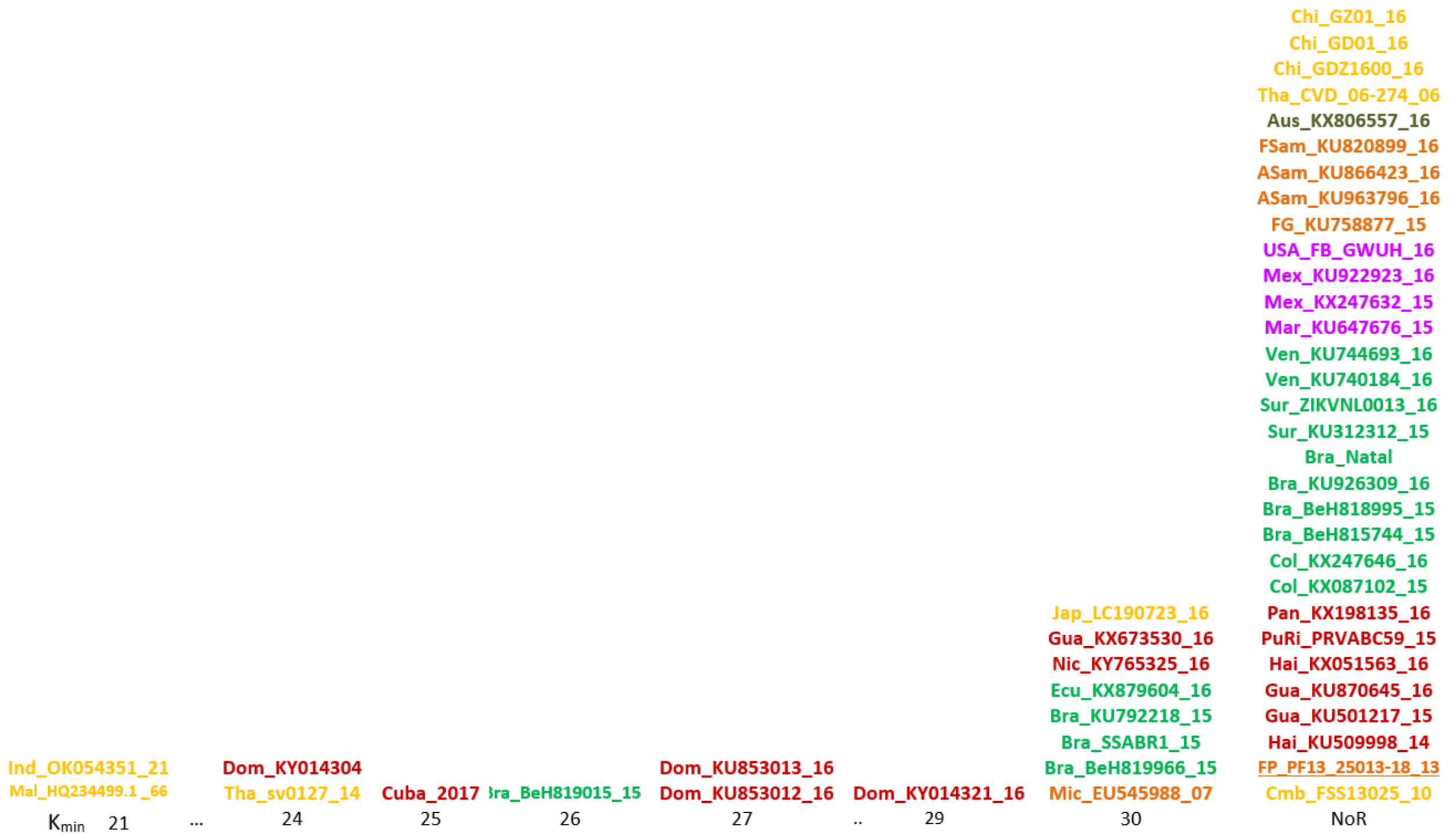

**Fig. 4.** Distribution of Asian strains of the Zika virus according to the $K_{\min}$ value, which defines the minimum replicator size generated from the original Hopfield neural network that was trained on KM-encoded sequences. Indian and Malaysian viruses are suggested to have earlier origin and highest complexity. Dominicana, Thailand and Cuban strains seem to arize later while most of the strains belong to the cell (3, NoR) and have low complexity and recent origin. Countries from near regions are shown in the same colors.

Table 9. Asian and American strains of Zika virus belonging to cell (3,M). The $K_{\min}$ values orrespond to the minimum size of the network processing KM-encoded genomes, and generating replicators are presented along with the corresponding sets of patterns treansmitted by replicators. The $K_{\max}$ values correspond to the size of replicator that has maximum number of transmitted patterns.

| Zika virus strain | Accession | $K_{\min}$ | Transmitted patterns at $K_{\min}$ | $K_{\max}$ | Transmitted patterns for $K_{\max}$ |
|---|---|---|---|---|---|
| Mal_HQ234499.1_66 | HQ234499.1 | 21 | 000000000000000001111 | 22 | 1111111111100000000000<br>1111111111000000000000<br>111111100000000000000<br>1111110000000000000000<br>1111100000000000000000<br>1111000000000000000000<br>1110000000000000000000 |
| Ind_OK054351_21 | OK054351 | 21 | 000000000000000001111 | 24 | 000000000000000011111111<br>000000000000000001111111<br>000000000000000000111111<br>000000000000000000011111<br>000000000000000000001111<br>000000000000000000000111 |
| Tha_sv0127_14 | MG770188.1 | 24 | 000000000000000000000111 | 30 | 000000000000000000000111111111<br>000000000000000000000011111111<br>000000000000000000000001111111 |
| Dom_KY014304_16 | KY014304.2 | 24 | 000000000000000000000111 | 26 | 00000000000000000000011111<br>00000000000000000000001111<br>00000000000000000000000111 |
| Cuba_2017 | MF438286 | 25 | 1110000000000000000000000 | 28 | 0000000000000000000001111111<br>0000000000000000000000001111 |
| Bra_BeH819015_15 | KU365778.1 | 26 | 00000000000000000000000111 | 29 | 00000000000000000000011111111<br>00000000000000000000000111111 |
| Dom_KU853012_16 | KU853012.1 | 27 | 000000000000000000000111111<br>000000000000000000000001111 | 27 | 000000000000000000000111111<br>000000000000000000000001111 |
| Dom_KU853013_16 | KU853013.1 | 27 | 000000000000000000000001111 | 30 | 000000000000000000000011111111<br>000000000000000000000001111111 |
| Dom_KY014321_16 | KY014321 | 29 | 00000000000000000000000011111 | 29 | 00000000000000000000000011111 |
| Ec_KX879604_16 | KX879604.1 | 30 | 000000111111111111111111000000<br>000000011111111111111111000000<br>000000001111111111111111100000<br>000000001111111111111111000000 | 30 | 000000111111111111111111000000<br>000000011111111111111111000000<br>000000001111111111111111100000<br>000000001111111111111111000000 |
| Nic_KY765325_16 | KY765325.1 | 30 | 000000011111111111111111000000<br>000000001111111111111111100000 | 30 | 000000011111111111111111000000<br>000000001111111111111111100000 |
| Gua_KX673530_16 | KX673530.1 | 30 | 000000000000000000000011111111<br>000000000000000000000001111111 | 30 | 000000000000000000000011111111<br>000000000000000000000001111111 |
| Bra_SSABR1_15 | KU707826.1 | 30 | 000000000000000000000001111111 | 30 | 000000000000000000000001111111 |
| Bra_BeH819966_15 | KU365779.1 | 30 | 000000000000000000000011111111 | 30 | 000000000000000000000011111111 |
| Bra_KU792218_15 | KU792218.1 | 30 | 000000000000000000000011111111 | 30 | 000000000000000000000011111111 |
| Jap_LC190723_16 | LC190723 | 30 | 111111111000000000000000000000<br>111111110000000000000000000000 | 30 | 111111111000000000000000000000<br>111111110000000000000000000000 |
| Mic_EU545988_07 | EU545988.1 | 30 | 000000001111111111111111100000<br>000000001111111111111111000000 | 30 | 000000001111111111111111100000<br>000000001111111111111111000000 |

Table 10. Asian strains of the Zika virus belonging to the cell (3, NoR) and presented in Fig.4.

| Country | Viral strain | Accession | Country | Viral strain | Accession |
|---|---|---|---|---|---|
| Cambodia | Cmb_FSS13025_10 | MH158236 | French-Polynesia | FP_PF13_251013-18_13 | KY766069.1 |
| Haiti | Hai_KU509998.3_14 | KU509998.3 | Guatemala | Gua_KU501217.1_15 | KU501217.1 |
| Haiti | Hai_KX051563_16 | KX051563.1 | Guatemala | Gua_KU870645.1_16 | KU870645.1 |
| Puerto-Rico | PuRi_PRVABC59_15 | MK713748 | Panama | Pan_KX198135_16 | KX198135.2 |
| Colombia | Col_KX087102_15 | KX087102.1 | Colombia | Col_KX247646_16 | KX247646.1 |
| Brazil | Bra_BeH815747_15 | KU365780 | Brazil | Bra_BeH818995_15 | KU365777.1 |
| Brazil | Bra_KU926309_16 | KU926309.2 | Brazil | Bra_Natal | KU527068.1 |
| Surinam | Sur_KU312312_15 | KU312312.1 | Surinam | Sur_ZIKVNL0013_16 | KU937936.1 |
| Venezuela | Ven_KU740184_16 | KU740184.2 | Venezuela | Ven_KU744693_16 | KU744693.1 |
| Martinique | Mar_KU647676_15 | KU647676.1 | Mexico | Mexico_KX247632_15 | KX247632.1 |
| Mexico | Mex_KU922923_16 | KU922923.1 | USA | US/FB_GWUH_16 | KU870645.1 |
| French Guiana | FG_KU758877_15 | KU758877.1 | American Samoa | ASam_KU963796_16 | KU963796.1 |
| American Samoa | ASam_KU866423_16 | KU866423.2 | Fiji Samoa | FSam_KU820899_16 | KU820899.2 |
| Australia | Aus_KX806557_16 | KX806557.3 | China | Chi_GDZ1600_16 | KU761564.1 |
| China | Chi_GD01_16 | KU740184 | China | Chia_GZ01_16 | KU820898.1 |
| Thailand | Tha_CVD_06-274_06 | MG645981.1 | | | |

## 3. Neural Replicator Analysis of Japanese encephalitis viral genomes

Five Japanese encephalitis virus (JEV) genotypes have been classified based on phylogenetic analysis of the viral envelope gene or the whole genome. It is assumed that JEV genotype 1 is the latest lineage and genotype 5 is the earliest known JEV lineage. Chronologically, this ancestral lineage has diverged to produce five virus genotypes in the sequence 5, 4, 3, 2 and 1, and the origin place of JEV is therefore Malaysia, Tibet and earlier − Indonesia. The NRA also consideres genotype 1 as recent, thereby separating this genotype as well as genotype 3 into 2 clusters each, with an interesting mirror atructure.

The distribution of JEV strains by the $K_{min}$ value, which determines the minimum sixe of the replicator generated from the original Hopfield neural network trained on KM-encoded sequences, is shown in Fig.5. It can be seen that strains belonging to genotype 3 form two well-separated groups. The first of them has a center at $K_{min}$ =21, and also partially located at $K_{min}$ =20 and $K_{min}$ =22. The second one presumably is located at $K_{min}$ =28. We will designate these two groups as 3A and 3B, respectively. It is noteworthy that all Indian strains belong to group 3A – this region is also home of the Indian Zika strain.

It is important to note that recently, in May 2023, International Committee of Taxonomy of Viruses revised the structure of the family *Flaviviridae,* deleting the genus *Flavivirus* and introducing a new genus *Orthoflavivirus* ("true" *Flavivirus*) instead of the genus *Flavivirus.* This wouldn't create any problems if the purpose be a eliminate some difficulties connected with similar names of the family *Flaviviridae* and the genus *Flavivirus*.

As stated in [11]: "The vernacular terms "flavivirus" are "flaviviral" are therefore ambiguous as it is unclear whether these words in various contexts refer to all members of the family or only to those of one of its genera. Because the family *Flaviviridae* is expected to expand considerably in the near future in part due to the discovery of numerous (he)pegiviruses and possibly the more distantly related Jīngmén tick-like viruses and other flavivirus-like entities, it is proposed to abolish the ambiguity by replacing the genus name "*Flavivirus*" with "*Orthoflavivirus*" (which roughly translates to "true flaviviruses" or "flaviruses sensu stricto"). But the following statement creates a problem:

> "Members of that genus would then be referred to as "orthoflaviviruses" – all orthoflaviviruses would be flaviviruses, but not all flaviviruses would be orthoflaviviruses" [11].

Indeed, when checking GenBank for the Japanese encephalitis viruses studied in this article, it can be found that about 40% of JEV still belong to the now defunct *Flavivirus* genus, and the remaining 60% (only) are now assigned to the *Orthoflavivirus* genus. Interestingly, all of the JEVs remaining in the *Flavivirus* genus originate from China. Fig.6 shows JEVs belonging to the new *Orthoflavivirus* genus, as well as JEVs which still are marked as belonging to the already removed *Flavivirus* genus. It is remarkable, that such groups as $K_{min}$ =22, 28, 29 become empty and now do not contain "*true*" *Flaviviruses*. On the other hand, the distribution of orthoflaviviruses retains a structure similar to that found in the original study (Fig. 6).

An interesting case is also associated with the appearance of JEV (KX945367) in Angola during the 2016 outbreak [12]. The patient, who also had the Yellow fever virus, did not travel abroad. Phylogenetic analysis showed that this strain of Japanese encephalitis virus belongs to genotype 3. Neural Repllicator Analysis showed a similar result, placing this virus at the center of subgroup 3A ($K_{min}$ =21), which contains more than two dozen strains shown in Fig.5. Moreover, the single replicator transimitting pattern of this Angola strain at the threshold value $K_{min}$ =21, (111110000000000000000), coincides with many similar replicator patterns of other JEV strains, such as Korea KV1899_99 (AY316157.1), Australia FU_95 (AF217620.1), China Heilongjang Ha3_60 (JN381872.1), India Vellore P20778_58 (AF080251.1) and China Fujan LYZ_57 (JN381869.1). The presence of the African JEV strain in subgroup 3A supports the suggestion that strains with lower $K_{min}$ values (we also assume that they have higher complexity) may emerge earlier than other.

On the other hand, most of the strains belonging the recent genotype 1 do not have replicators for KM-encoded genome (at least for K≤30) and, therefore are located mainly in the cell (3, NoR) forming subgroup 1A, but are also partially located in another small subgroup 1B centered at $K_{min}$ =24 (Fig.5). Qualitatively, this is also consistent with the assumption that recent genotype 1 strains have low complexity and high $K_{min}$ values. But here a certain contradiction arises: strains of genotypes 4 (Indonesia) and 5 (Tibet, Malaysia) are also assigned by the NRA to the recent subgroup 1A. Remarkably, as noted in [13], two strains of genotype 5, "strain XZ0934 (Tibet) isolated in 2009, which was not included in earlier analyzes, showed a significant difference from Muar (Malaysia) isolate [14]". In contrast, NRA reliably grouped these two strains into subgroup 1A. This is not surprising for a reemerged Tibetian strain, but contradicts the conclusion made in varioust studies that genotype 5 is the earliest and that also Tibet or Malaysia may be the origin geographic regions of JEV. On the other hand, NRA replaces this location of JEV origin with India, which is fairly close to Tibet.

Thus, NRA clearly indicates that, as in the case of Zika virus, the JEV can also be spread to Asia from India and have its origin in India or Africa. Undoubtedly, this conclusion needs further research and substantiation.

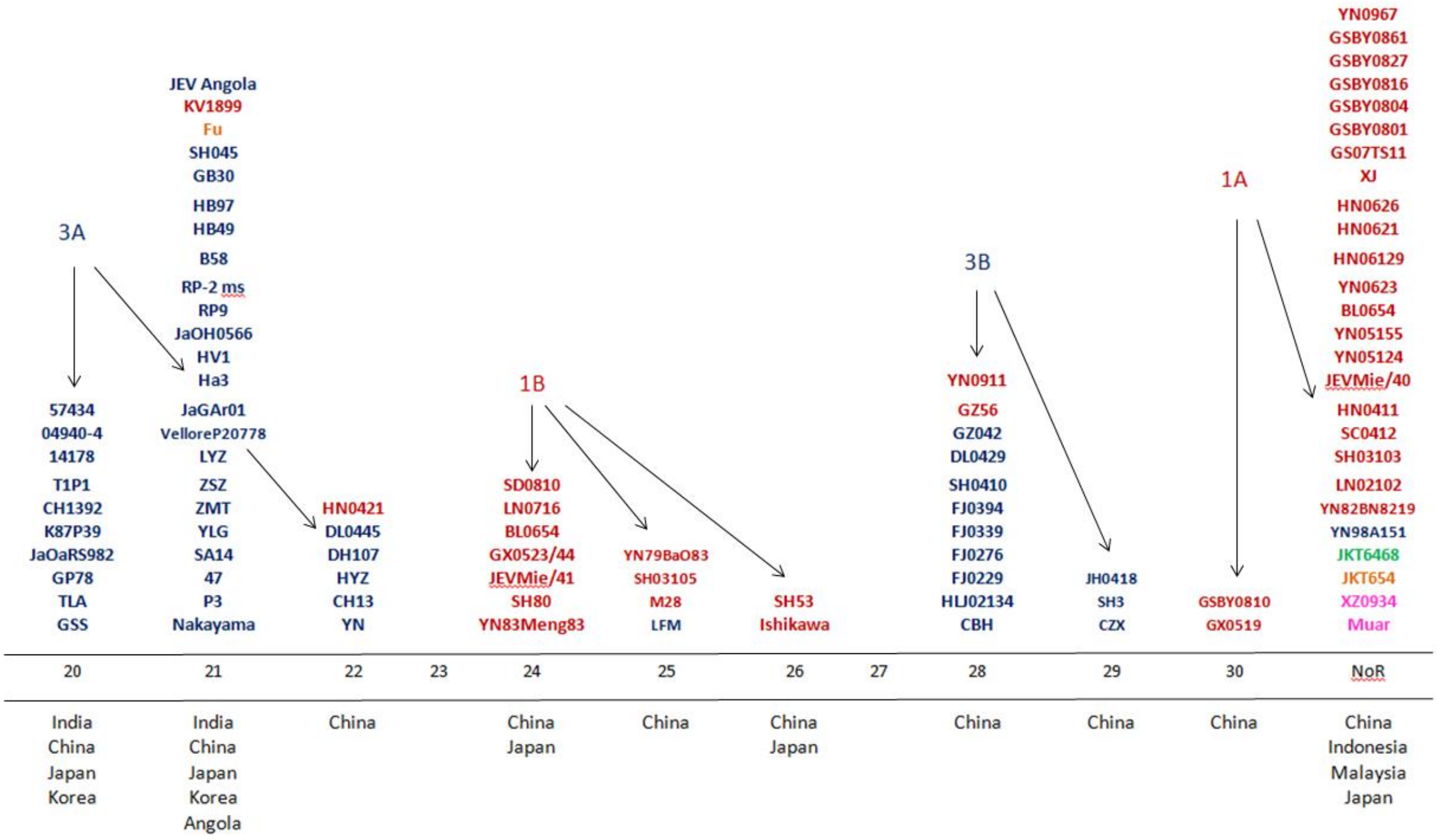

**Fig. 5.** Distribution of Japanese encephalitis virus strains according to the $K_{min}$ value, which defines the minimum replicator size generated from the original Hopfield neural network that was trained on KM-encoded sequences. Genotype 1 strains are shown in red, genotype 2 in orange, genotype 3 in blue, genotype 4 in green, and genotype 5 in purple.

Table 11. Japanese encephalitis viruses presented in Figs. 5,6

| Country Province | Viral strain | Ortho/Flavi May 28 2023 | Year | Accession | Genotype | NRA subgroup before May2023 | $K_{\min}$ |
|---|---|---|---|---|---|---|---|
| China Beijing | GSS | Orthoflavivirus | 1960s | JF706275 | 3 | 3A | 20 |
| China Liaoning | TLA | Flavivirus | 1971 | JN381868 | 3 | | |
| India | GP78 | Orthoflavivirus | 1978 | AF075723 | 3 | | |
| Japan | JaOaRS982 | Orthoflavivirus | 1982 | M18370 | 3 | | |
| South Korea | K87P39 | Orthoflavivirus | 1987 | AY585242 | 3 | | |
| China Taiwan | CH1392 | Orthoflavivirus | 1990 | AF254452 | 3 | | |
| China Taiwan | T1P1 | Orthoflavivirus | 1997 | AF254453 | 3 | | |
| India | 14178 | Orthoflavivirus | 2001 | EF623987 | 3 | | |
| India | 04940-4 | Orthoflavivirus | 2002 | EF623989 | 3 | | |
| India | 57434 | Orthoflavivirus | 2005 | EF623988 | 3 | | |
| Japan | Nakayama | Orthoflavivirus | 1935 | EF571853 | 3 | | 21 |
| China Beijing | P3 | Orthoflavivirus | 1949 | U47032 | 3 | | |
| China Heilongjiang | 47 | Orthoflavivirus | 1950s | JF706269 | 3 | | |
| China | SA14-14-2 | Orthoflavivirus | 1954 | KX254417 | 3 | | |
| China Fujian | YLG | Orthoflavivirus | 1955 | JF706280 | 3 | | |
| China Fujian | ZMT | Orthoflavivirus | 1955 | JF706283 | 3 | | |
| China Fujian | ZSZ | Flavivirus | 1955 | JN381862 | 3 | | |
| China Fujian | LYZ | Flavivirus | 1957 | JN381869 | 3 | | |
| India | Vellore P20778 | Orthoflavivirus | 1958 | AF080251 | 3 | | |
| Japan | JaGAr 01 | Orthoflavivirus | 1959 | AF069076 | 3 | | |
| China Heilongjiang | Ha3 | Flavivirus | 1960s | JN381872 | 3 | | |
| China Taiwan | HVI | Orthoflavivirus | 1965 | AF098735 | 3 | | |
| Japan | JaOH0566 | Orthoflavivirus | 1966 | AY508813 | 3 | | |
| China Taiwan | RP-2 ms | Orthoflavivirus | 1985 | AF014160 | 3 | | |
| China Taiwan | RP9 | Orthoflavivirus | 1985 | AF014161 | 3 | | |
| China Yunnan | B58 | Orthoflavivirus | 1989 | FJ185036 | 3 | | |
| China Yunnan | HB49 | Orthoflavivirus | 1990 | JF706284 | 3 | | |
| China Yunnan | HB97 | Orthoflavivirus | 1990 | JF706285 | 3 | | |
| Australia | FU | Orthoflavivirus | 1995 | AF217620 | 2 | | |
| China Yunnan | GB30 | Orthoflavivirus | 1997 | FJ185037 | 3 | | |
| Korea | KV1899 | Orthoflavivirus | 1999 | AY316157 | 1 | | |
| China Shanghai | SH04-5 | Flavivirus | 2004 | JN381866 | 3 | | |
| Angola | JEV Angola | Orthoflavivirus | 2016 | KX945367 | 3 | | |
| China Yunnan | YN | Flavivirus | 1954 | JN381871 | 3 | | 22 |
| China Sichuan | CH13 | Flavivirus | 1957 | JN381870 | 3 | | |
| China Yunnan | HYZ | Flavivirus | 1979 | JN381853 | 3 | | |
| China Yunnan | DH107 | Flavivirus | 1989 | JN381873 | 3 | | |
| China Yunnan | DL04-45 | Flavivirus | 2004 | JN381854 | 3 | | |
| China Henan | HN04-21 | Flavivirus | 2004 | JN381841 | 1 | | |
| China Yunnan | YN83-Meng83-54 | Orthoflavivirus | 1983 | JF706282 | 1 | 1B | 24 |
| China Shanghai | SH-80 | Flavivirus | 2001 | JN381848 | 1 | | |
| Japan | JEV/sw/Mie/41 | Orthoflavivirus | 2002 | AB241119 | 1 | | |
| China Guangxi | GX0523/44 | Flavivirus | 2005 | JN381832 | 1 | | |
| China Guangxi | BL06-54 | Orthoflavivirus | 2006 | JF706271 | 1 | | |
| China Liaoning | LN0716 | Flavivirus | 2007 | JN381849 | 1 | | |
| China Shandong | SD0810 | Orthoflavivirus | 2008 | JF706286 | 1 | | |
| China Fujian | LFM | Flavivirus | 1955 | JN381863 | 3 | | 25 |
| China Yunnan | M28 | Orthoflavivirus | 1977 | JF706279 | 1 | | |
| China Yunnan | YN79-Bao83 | Flavivirus | 1979 | JN381851 | 1 | | |
| China Shanghai | SH03-105 | Flavivirus | 2003 | JN381846 | 1 | | |
| Japan | Ishikawa | Orthoflavivirus | 1994 | AB051292 | 1 | | 26 |
| China Shanghai | SH-53 | Flavivirus | 2001 | JN381850 | 1 | | |
| China Fujian | CBH | Flavivirus | 1954 | JN381860 | 3 | 3B | 28 |
| China Heilongjiang | HLJ02-134 | Orthoflavivirus | 2002 | JF706276 | 3 | | |
| China Fujian | FJ02-29 | Orthoflavivirus | 2002 | JF706273 | 3 | | |
| China Fujian | FJ02-76 | Flavivirus | 2002 | JN381867 | 3 | | |
| China Fujian | FJ02-39 | Flavivirus | 2003 | JN381859 | 3 | | |
| China Fujian | FJ02-94 | Flavivirus | 2003 | JN381858 | 3 | | |
| China Yunnan | DL04-29 | Orthoflavivirus | 2004 | JF706272 | 3 | | |
| China Shanghai | SH04-10 | Flavivirus | 2004 | JN381856 | 3 | | |
| China Guizhou | GZ04-2 | Flavivirus | 2004 | JN381857 | 3 | | |
| China Guizhou | GZ56 | Orthoflavivirus | 2006 | HM366552 | 1 | | |
| China Yunnan | YN0911 | Orthoflavivirus | 2009 | JF706267 | 1 | | |
| China Fujian | CZX | Flavivirus | 1954 | JN381857 | 3 | | 29 |
| China Shanghai | SH-3 | Flavivirus | 1987 | JN381864 | 3 | | |
| China Yunnan | JH04-18 | Flavivirus | 2004 | JN381855 | 3 | | |
| China Guangxi | GX0519 | Flavivirus | 2005 | JN381835 | 1 | | 30 |
| China Gansu | GSBY0810 | Flavivirus | 2008 | JN381840 | 1 | | |
| Malaysia | Muar | Orthoflavivirus | 1952 | HM596272 | 5 | | |
| Indonesia Java | JKT654 | Orthoflavivirus | 1978 | HQ223287 | 2 | | |
| Indonesia | JKT6468 | Orthoflavivirus | 1981 | AY184212 | 1 | | |
| China Yunnan | YN82-BN8219 | Flavivirus | 1982 | JN381834 | 1 | | |
| China Liaoning | LN02-102 | Orthoflavivirus | 2002 | JF706278 | 1 | | |
| China Shanghai | SH03-103 | Flavivirus | 2003 | JN381847 | 1 | | |
| China Yunnan | YN98-A151 | Flavivirus | 2003 | JN381861 | 3 | | |
| China Sichuan | SC04-12 | Flavivirus | 2004 | JN381839 | 1 | | |
| China Henan | HN04-11 | Flavivirus | 2004 | JN381831 | 1 | 1A | NoR |
| Japan | JEV/sw/Mie/40/ | Orthoflavivirus | 2004 | AB241118 | 1 | | |
| China Yunnan | YN05124 | Orthoflavivirus | 2005 | JF706281 | 1 | | |
| China Henan | HN06129 | Orthoflavivirus | 2006 | JF706277 | 1 | | |
| China Henan | HN0621 | Flavivirus | 2006 | JN381830 | 1 | | |
| China Henan | HN0626 | Flavivirus | 2006 | JN381837 | 1 | | |
| China Yunnan | YN0623 | Flavivirus | 2006 | JN381836 | 1 | | |
| China Guangxi | BL06-54 | Orthoflavivirus | 2006 | JF706271 | 1 | | |
| China | XJ69 | Orthoflavivirus | 2007 | EU880214 | 1 | | |
| China Gansu | GS07-TS11 | Flavivirus | 2007 | JN381843 | 1 | | |
| China Gansu | GSBY0810 | Flavivirus | 2008 | JN381840 | 1 | | |
| China Gansu | GSBY0804 | Flavivirus | 2008 | JN381844 | 1 | | |
| China Gansu | GSBY0816 | Flavivirus | 2008 | JN381842 | 1 | | |
| China Gansu | GSBY0827 | Flavivirus | 2008 | JN381845 | 1 | | |
| China Gansu | GSBY0861 | Flavivirus | 2008 | JN381833 | 1 | | |
| China Yunnan | YN0967 | Orthoflavivirus | 2009 | JF706268 | 1 | | |
| China Tibet | XZ0934 | Orthoflavivirus | 2009 | JF915894 | 5 | | |

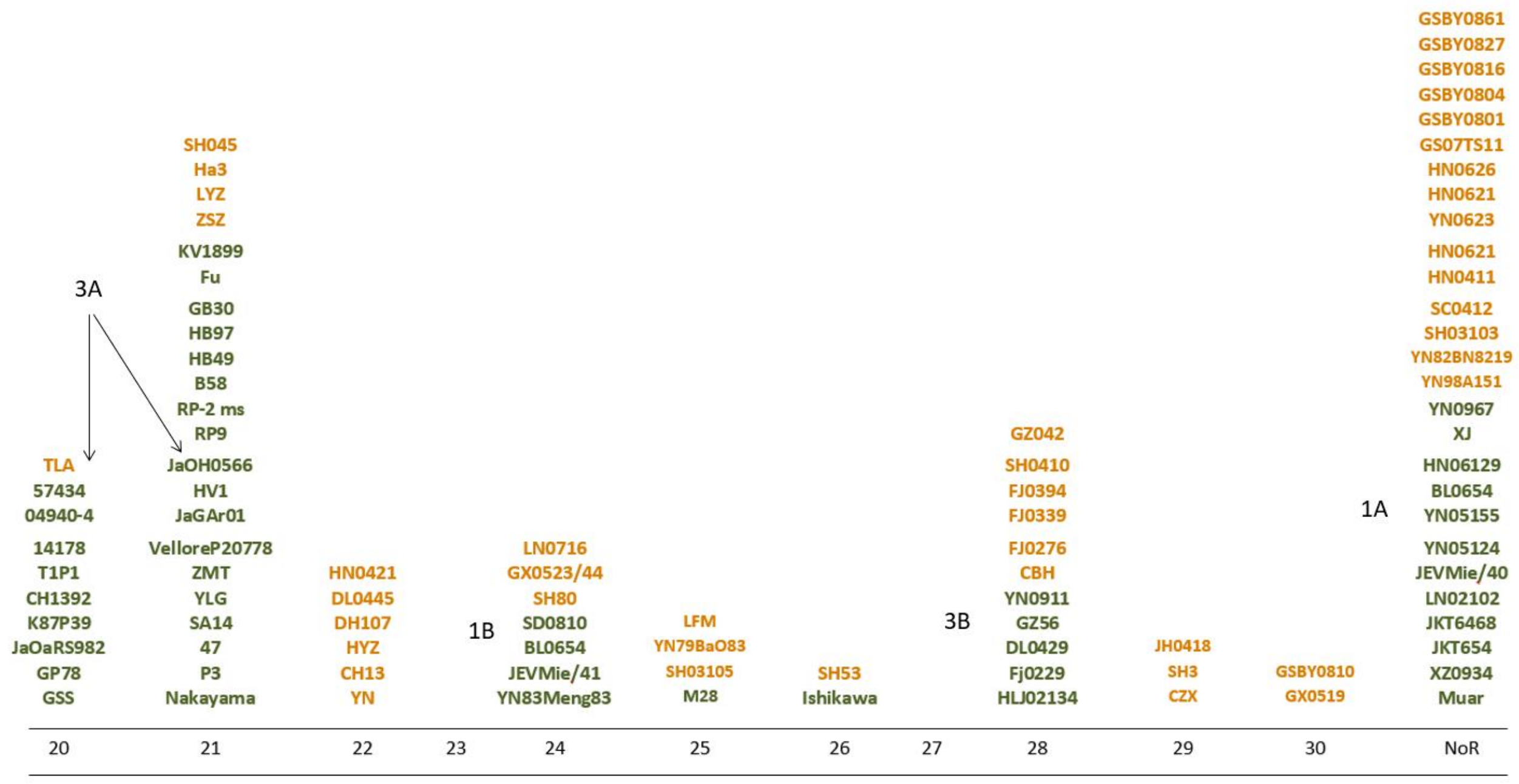

**Fig. 6.** Distribution of Japanese encephalitis virus strains according to the $K_{min}$ value, which defines the minimum replicator size generated from the original Hopfield neural network that was trained on KM-encoded sequences. Strains renamed as belonging to the genus *Orthoflavivirus* are shown in green, while those remaining in the canceled genus *Flavivirus* are shown in yellow. It can be seen that the distribution of orthoflusviviruses generally retains the structure shown in Fig. 5 (subgroups 1A, 1B, 3A, 3B), with narrowed subgroups.

## Conclusion

We have demonstrated that the Neural Replicator Analysis of viral genomes of the *Flavivirus* genus filled in some new cells in the joint viral genome table presented earlier in [4] (Fig. 7). In particular, new important cell (9, NoR) is filled with tick-borne flaviviruses, while the other cells are populated with presumably mosquito-borne viruses. While it was demonstrated in [4] that NRA can place in the same cells the genomes of viruses that have the same *hosts* and cause similar *diseases*, here we show with even stronger arguments that NRA can place viral genomes in *vector-specific* cells.

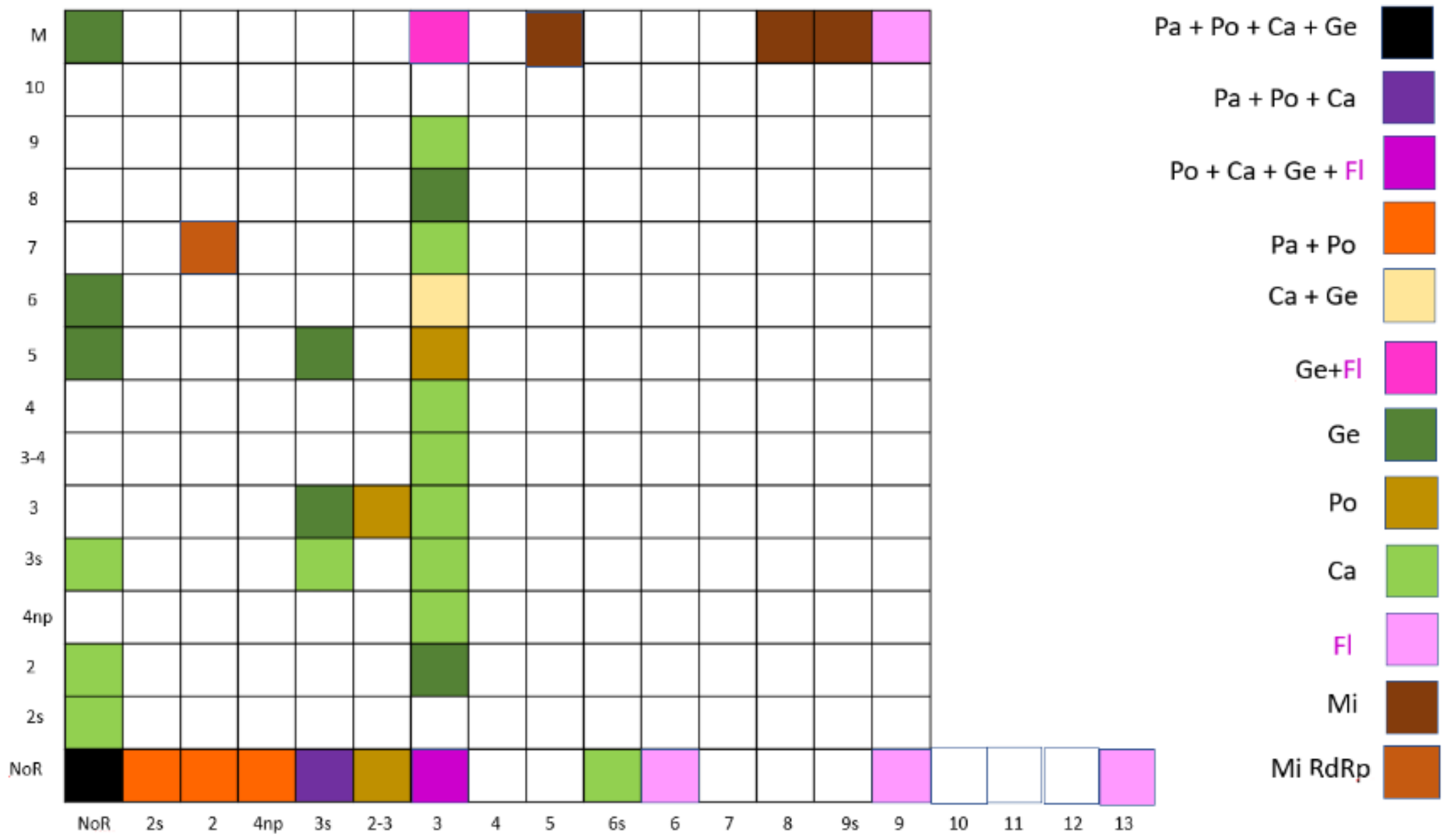

**Fig. 7.** Joint table of viral genomes showing cells populated according to NRA data obtained for the genus *Flavivirus* (Fl) and also presented in [4] families *Papillomaviridae* (human) – Pa, *Polyomaviridae* – Po, *Caulimoviridae* – (Ca), *Geminiviridae* – Ge and also *Mitoviridae* – Mi (Gigaspora margarita mitovirus 1, NC_040702.1 – (5, M), Cronartium ribicola mitovirus 5, NC_030399.1 –(8, M); Fusarium poae mitovirus 4, NC_030864.1 – (9s, M) ; Rhizoctonia mitovirus 1 RdRp, NC_040563.1 – (2, 7) ). Some cells are filled with genomes of viruses belonging to different families.

In addition, the strains of the Japanese encephalitis virus, as well as the strains of the Zika virus studied in this article, occupy not one, but two common cells of the table. The results of the NRA of Zika viral strains suggest that the earliest strain in Asia is an Indian strain that spread from Africa (Uganda) to the East. Examination of the fine structure of Japanese encephalitis virus strain sets shows that their generally accepted genotypes 1 and 3 can be clearly divided into two well-defined subgenotypes. We also argue that probably Indian strains of this virus can also be considered the earliest known Asian strains that subsequently evolved and spread to other Asian countries. Some speculation has also been made about the association of the temporal direction of virus evolution with the reduction in genome complexity, as well as the dependence of higher complexity on the characteristics of the Replicator Tables, especially on the smaller network size in which replicators arise when the KM-encoded genome is analyzed. Here, a higher complexity is associated with the non-ordered, but random nature of the KM-encoded genome. Indeed, when studying the generation of replicators by a germ network with a randomly chosen connections weights, it was found that replicator patterns do indeed have a monotonically decreasing (and non-periodic) activity [15,16]. Of course, these relationships of viral evolution, complexity, and nature of replicator patterns require further critical analysis. However, we hope that the current study of Neuronal Replicator Analysis of flaviviruses shows that this method can produce interesting and non-trivial results.